\documentclass[12pt]{article} 
\usepackage[tablesfirst,notablist,nomarkers]{endfloat} 
\usepackage{ol2}

\usepackage[draft]{hyperref}
\usepackage{amsmath}
\usepackage{amssymb}
\usepackage{graphicx}

\begin{document}


\title{Octave spanning spectra and pulse compression by nondegenerate cascaded four-wave mixing: 2D simulation and experiment}


\author{Jo\~{a}o L. Silva,$^{1}$ Rosa Weigand,$^{2}$ Helder M. Crespo,$^{1,*}$ }

\address{
$^1$IFIMUP/IN - Institute of Nanoscience and Nanotechnology and Departamento de F\'{i}sica, Faculdade de Ci\^{e}ncias, Universidade do Porto, Rua do Campo Alegre, 687, 4169-007 Porto, Portugal \\
$^2$Departamento de \'{O}ptica, Facultad de Ciencias F\'{i}sicas, Universidad Complutense de Madrid, Ciudad Universitaria s/n, 28040 Madrid, Spain \\
$^*$Corresponding author: hcrespo@fc.up.pt
}

\begin{abstract}We present a theoretical model and respective numerical simulation of cascaded four-wave mixing of femtosecond pulses in bulk $\chi^{\left(3\right)}$ media, evidencing the importance of two-dimensional interaction geometries in the efficient generation of frequency-converted beams, and in agreement with the experimental results in terms spectral and spatial distribution and efficiency, demonstrating compression without the need of complex amplitude or phase control.
\end{abstract}



\noindent The simultaneous phase-matching of multiple nonlinear optical processes can give rise to interesting cascading phenomena, with each consecutive process efficiently making use of optical signals generated by the previous ones. In particular, cascaded second-order processes \cite{Saltiel2005} based on sum- and difference-frequency generation of intense laser pulses in $\chi^{\left(2\right)}$ media can be very useful for generating new frequencies \cite{Zhang2004} beyond the usual signal and idler waves produced in single-step processes. However, these can only occur in non-centrosymmetric materials.
Third-order nonlinear processes, namely four-wave mixing (FWM), have the advantage that $\chi^{\left(3\right)}$ is nonzero for all media, without any restrictions imposed by symmetry. Therefore, FWM can occur in any material, including isotropic media such as glasses, gases and plasmas. In particular, nondegenerate FWM of noncollinear laser pulses with frequencies $\omega_{0}$ and $\omega_{1}$ ($\omega_{1}>\omega_{0}$) can be geometrically phase-matched in media with positive dispersion to generate a third order difference-frequency signal $\omega_{2}=2\omega_{1}-\omega_{0}$\cite{Auston1971}. Cascading of this process (generating pulses at $\omega_{n}=n\omega_{1}-(n-1)\omega_{0}$ where $n$ is the beam number or order, with negative values denoting frequency-downshifted beams) would seem at first impossible due to the need of different phase-matching angles for each intermediate step. However, this condition can be greatly relaxed by using broad bandwidth pump pulses.

In the first demonstration of highly nondegenerate cascaded four-wave mixing (CFWM) of femtosecond pulses in bulk $\chi^{\left(3\right)}$ media \cite{Crespo2000} it was shown that two noncollinear pump pulses with different frequencies overlapping in a nonresonant Kerr medium can generate a series of angularly separated multicolored pulses with broad bandwidths and frequencies extending from the infrared to the ultraviolet. Broadband phase-matching of the corresponding multiple processes was achieved by adjusting the interaction angle between the pumps, and the use of a thin slide as the nonlinear medium minimized material dispersion and competing nonlinear optical effects such as self- and cross-phase modulation while maintaining high conversion efficiencies (approx. 10\%). 
The efficient extension of CFWM generation down to $160\ {\rm nm}$ in gas-filled hollow waveguides \cite{Misoguti2001}, with phase-matching achieved by tuning the gas pressure, further evidenced the universal nature of nonresonant CFWM processes. More recently, CFWM was also observed in noncollinear four-wave parametric amplification of ultrashort laser pulses in water \cite{Dubietis2007}.
Another interesting characteristic of nonresonant FWM and CFWM processes is the possibility of generating and synthesizing compressed pulses. This is again a direct consequence of the modulation of the incident fields by the material excitation, in its turn proportional to the product of the initial pump field envelopes. Fuji et al. obtained ultraviolet $266\ {\rm nm}$ pulses with $12\ {\rm fs}$ in duration by FWM of the fundamental and second harmonic pulses from a $25\ {\rm fs}$ Ti:sapphire laser in a gas filament, where cascaded generation of radiation at $200\ {\rm nm}$ was also observed \cite{Fuji2007}.
In recent work, polarization gated cross-correlation frequency-resolved optical gating (PG-XFROG) measurements of several simultaneous cascaded beams generated in fused silica and covering 1.5-octaves in bandwidth were performed that support the possibility of using CFWM to Fourier-synthesize high-power optical pulses in the single-cycle regime \cite{Crespo2009,Weigand2009}.

The interest and characteristics of CFWM processes have been calling for theoretical and numerical models capable of accurately describing its detailed features and potential. Hart et al. \cite{Hart1994} used coupled wave equations as well as the one dimensional nonlinear Schr\"{o}dinger equation to describe multiple FWM of nanosecond pulses in optical fibers. Lichtman et al. \cite{Lichtman1987} obtained exact analytical solutions of the total field for 1D propagation of 2 or 3 monochromatic input waves but assuming a nondispersive medium and perfect phase matching.

We found out that in order to accurately reproduce the main features of broadband CFWM and to account for geometrical phase matching, a two-dimensional approach is required. In this letter we present a numerical study of CFWM of noncollinear femtosecond pulses with Gaussian spatial profiles crossing in a thin fused silica slide. In our model we consider the propagation of the pulse envelope on the 2D plane defined by the pump beams, using a reference frame that moves along the propagation axis $z$ with the group velocity of one of the pumps, as given by Eqs. \ref{eq:lin}-\ref{eq:nlin}.

\begin{align}\label{eq:lin}
\frac{\partial A\left(x,\omega\right)}{\partial z} & = & \frac{i}{2k_{0}^{'}\left(\omega\right)}\left[\frac{\partial^{2}}{\partial x^{2}}+k^{2}\left(\omega\right)-k_{0}^{2} \right. \nonumber \\
 & & \left. -\frac{2\delta\omega k_{0}}{v_{g}}-\frac{\delta\omega^{2}}{v_{g}^{2}}\right]A
\end{align}

\begin{align}\label{eq:nlin}
\frac{\partial A\left(x,t\right)}{\partial z} = & \frac{3\chi^{\left(3\right)}}{8ik_{0}c^{2}}\left[-\omega_{0}^{2}\left|A\right|^{2}A \right. & \nonumber \\
 & \left. -2i\omega_{0}\frac{\partial\left(\left|A\right|^{2}A\right)}{\partial t} +\frac{\partial^{2}\left(\left|A\right|^{2}A\right)}{\partial t^{2}}\right] &
\end{align}
where $A$ is the envelope of the pulse with electric field $E\left(x,t\right)=A\left(x,t\right)\exp\left(ik_{0}z-i\omega_{0}t\right)$, $\delta\omega=\omega-\omega_{0}$, $v_{g}$ is the group velocity at the reference frequency $\omega_{0}$, $k\left(\omega\right)=n_{\omega}\omega/c$, $k_{0}=k\left(\omega_{0}\right)$ and $k_{0}^{'}\left(\omega\right)=k_{0}+\delta\omega/v_{g}$. Eq. \ref{eq:lin} describes linear propagation in the frequency domain, and includes diffraction and dispersion (although due to the small longitudinal dimension of the medium the group velocity mismatch and pump walk-off are practically negligible). Eq. \ref{eq:nlin} contains the nonlinear terms in the time domain, which comprise the instantaneous nonresonant Kerr effect and self-steepening. The accurate evaluation of the self-steepening terms is essential because the generated spectral width is of the same order as $\omega_{0}$. No additional delayed response terms were required to faithfully reproduce the experimental results.

\begin{figure}[htb]
\centerline{
\includegraphics[width=8.3cm]{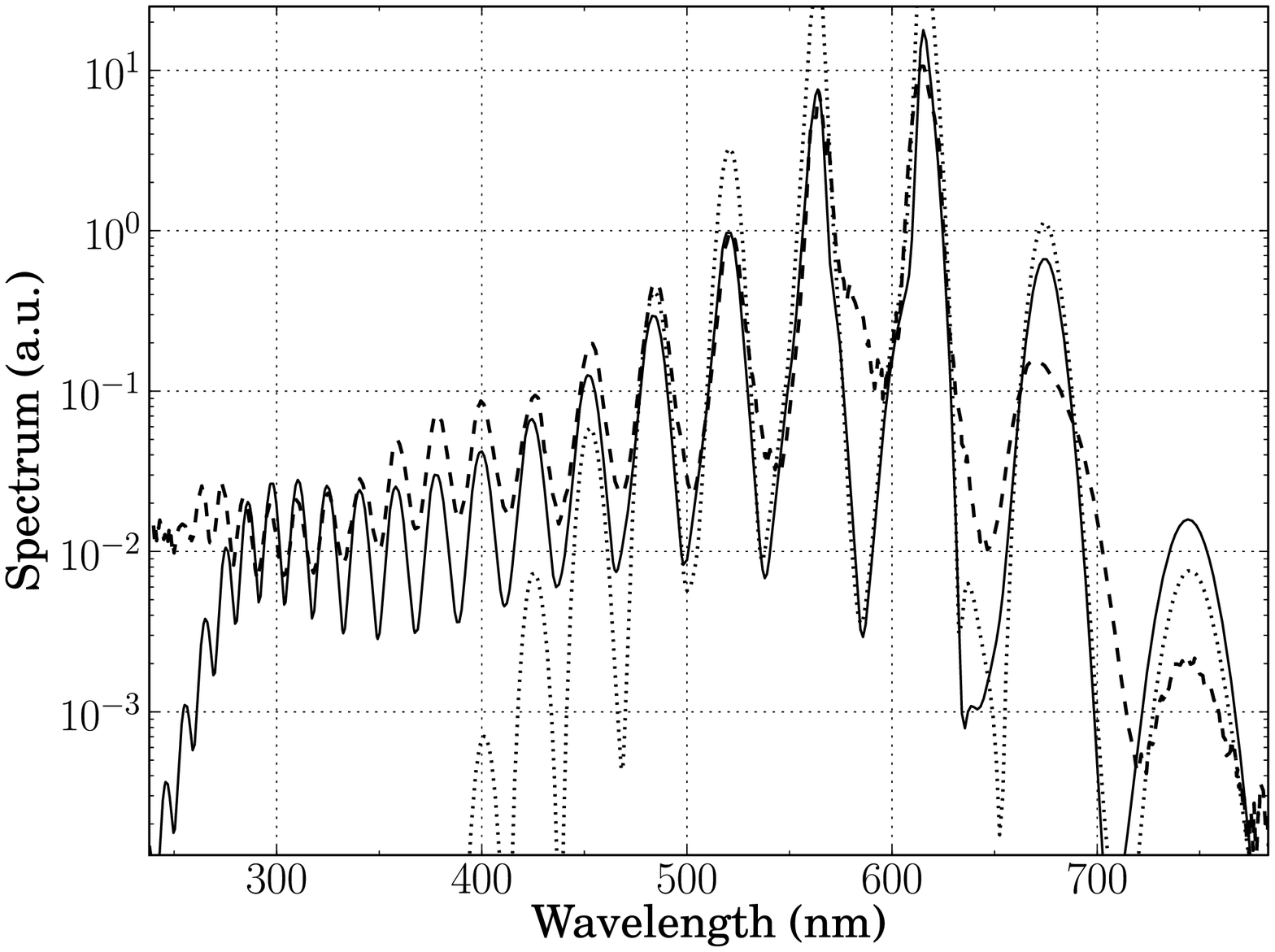}}
 \caption{\label{fig:spectra} Spectrum of the multiple CFWM orders generated in fused silica. The spectrum spans over 1.5 octaves (the two pump frequencies correspond to the two larger peaks). Dashed line: Experimental results (see \cite{Weigand2009}), Solid line: Numerical simulation, Dotted line: Numerical simulation for collinear pumps.
}
\end{figure}

We solve these equations in a split step fashion, the first using a (2,2)-Pad\'{e} approximant for wide-angle propagation  as two (1,1) steps\cite{Hadley1992a}, and the second using a second order Runge-Kutta method, with time derivatives approximated by 5-point finite differences. We used the appropriate parameters to simulate our experiment described in \cite{Weigand2009}. The green ($565\ {\rm nm}$, $60\ {\rm fs}$, $1.5\times 10^{12}\ {\rm W/cm^2}$) and orange ($614.8\ {\rm nm}$, $80\ {\rm fs}$, $1.8\times 10^{12}\ {\rm W/cm^2}$) transformed-limited pump pulses propagate in $172\ {\rm \mu m}$ of fused silica (corresponding to a $150\ {\rm \mu m}$ thick slide tilted at $45$ degrees, which reduces the angle between the normal to the slide and the generated frequency upshifted beams thus preventing total internal reflection at the silica-air interface), crossing at an internal angle of $1.57$ degrees. The green pulse is also delayed $48\ {\rm fs}$ with respect to the orange at the entrance of the slide, which resulted in better defined spectra for the CFWM orders, in agreement with the experimental measurements.

The simulations provide the time dependent field distribution at the end of the slide as a function of the transverse coordinate $x$ and frequency $\omega$. The corresponding spectrum (obtained by integration along the transverse direction) is given in Fig. \ref{fig:spectra}, clearly showing the asymmetric generation of up and down-converted orders. The efficiency of the generated frequency upshifted pulses follows the same pattern as the experiment, although it has a sharper cutoff after the 12th order that depends on the interaction angle and the pulse energy. This discrepancy is possibly due to experimental calibration errors which are larger at the spectral edges. For collinear pumps the efficiency is low because the process is not phase matched. As the angle is increased ($\gtrsim 1.4$ degrees) phase matching favors the generation of higher order modes, as shown in \cite{Weigand2009}. At larger internal angles ($>2$ degrees) the overall efficiency drops rapidly.

\begin{figure}[htb]
\centerline{
\includegraphics[width=8.3cm]{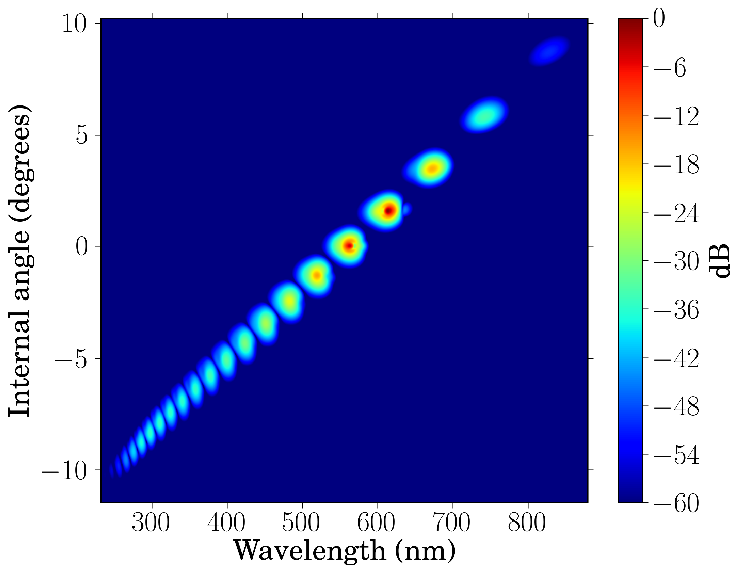}}
 \caption{\label{fig:theta_lambda} (Color online) Simulated $\theta-\lambda$ spectrum of the CFWM beams, clipped at $10^{-6}$ of the maximum.}
\end{figure}

The $\theta-\lambda$ spectrum (Fig. \ref{fig:theta_lambda}) shows the main features of the CFWM process in the corresponding domains. The various orders are emitted with different wavelengths and angles, with decreasing angular separation between beams of increasing order.

\begin{figure}[htb]
\centerline{
\includegraphics[width=8.3cm]{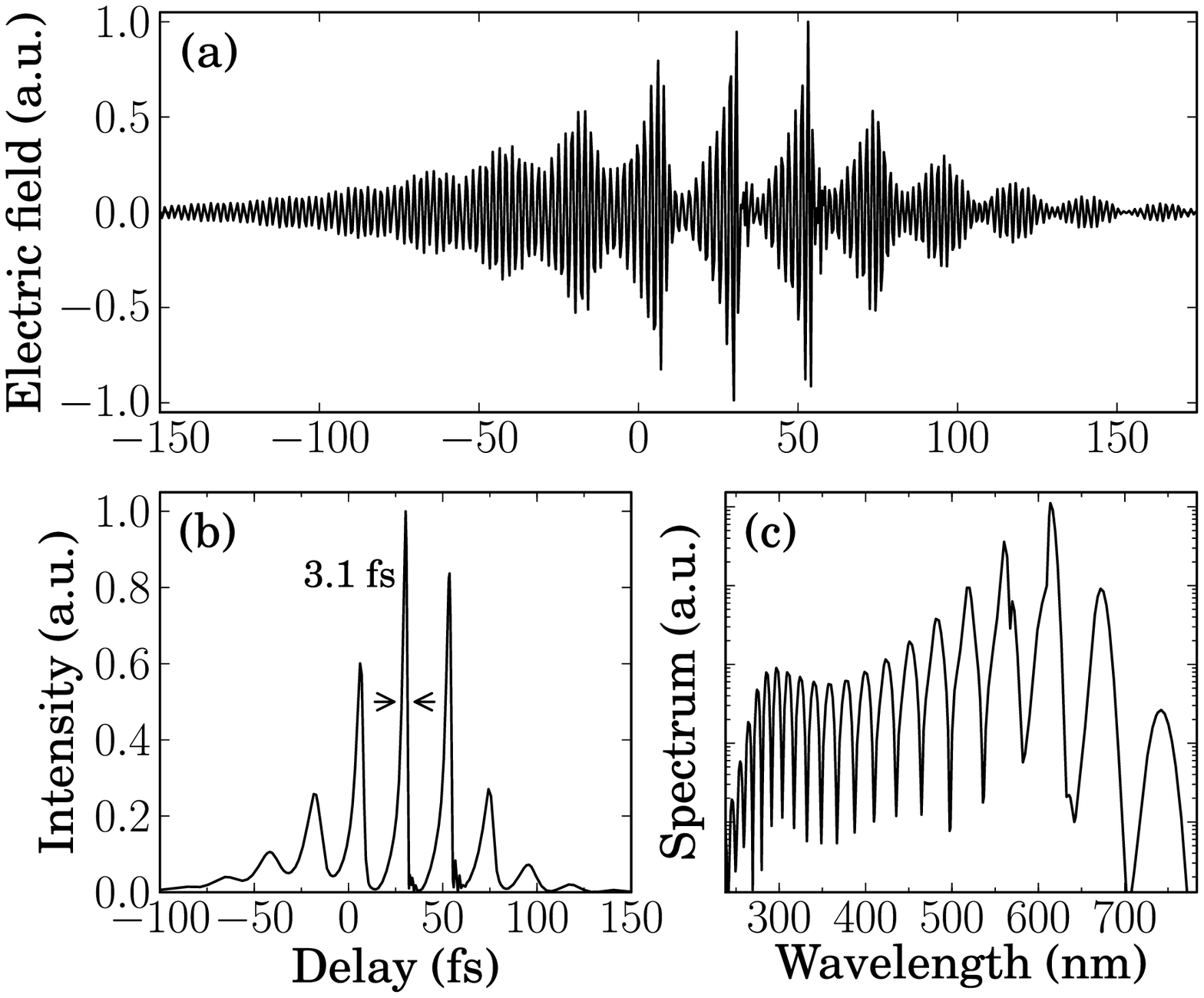}}
 \caption{\label{fig:time} Recombined and focused CFWM pulses. (a) Electric field as a function of the delay (in ${\rm fs}$). (b) Intensity. (c) Spectrum (logarithmic scale).}
\end{figure}

The simultaneous spatial and temporally superimposed CFWM orders have been characterized experimentally with PG-XFROG where a portion of the orange pump pulse acted gate \cite{Crespo2009,Weigand2009}. Here, to simulate the focusing and obtain the total electric field needed to calculate the PG-XFROG traces we simply subtracted the linear phase (group delay) acquired due to the transverse propagation through the fused silica slide. The resulting field intensity and respective spectrum are shown on Fig. \ref{fig:time} which shows a train compressed pulses separated by the pump beat period ($23\ {\rm fs}$) as expected. The central peak has a duration of $3.1\ {\rm fs}$ FWHM.

\begin{figure}[htb]
\centerline{
\includegraphics[width=8.3cm]{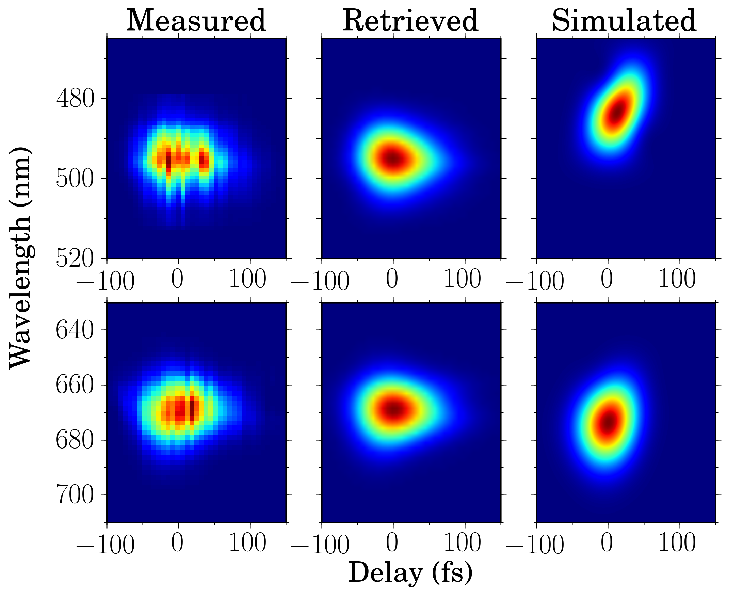}}
 \caption{\label{fig:xfrog} (Color online) Experimental (measured and retrieved) and simulated PG-XFROG traces of the second frequency-upshifted pulse (top row) and first downshifted pulse (bottom row). The gate pulse is the $80\ {\rm fs}$ input orange pump. The retrieved pulse durations are $30.7$ and $40.2\ {\rm fs}$ for second upshift and first downshift respectively (FROG error was $0.014$ and $0.008$ for a 128x128 grid).}
\end{figure}

\begin{figure}[htb]
\centerline{
\includegraphics[width=8.3cm]{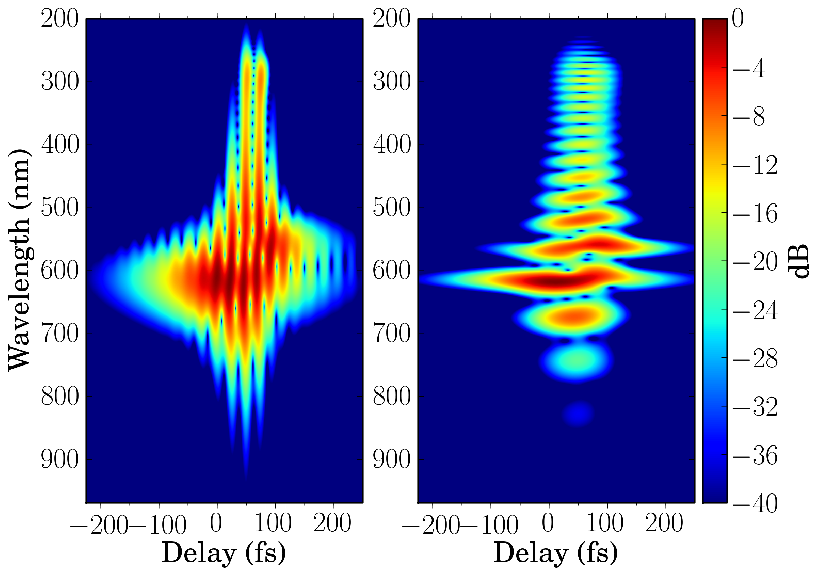}}
 \caption{\label{fig:xfrog_all} (Color online) Simulated PG-XFROG, clipped at $10^{-4}$ of the maximum, for the recombined and focused CFWM field assuming a $20\ {\rm fs}$ gate pulse. Right: $80\ {\rm fs}$ gate pulse.}
\end{figure}

Fig. \ref{fig:xfrog} compares the experimental and simulated PG-XFROG traces for the second frequency-upshifted pulse and the first downshifted pulse. We see that the simulation results are in good agreement with the measurements, providing similar values for the bandwidth and duration of the generated pulses (which are also shorter than the pumps). The observed difference in the central frequency of the upshifted pulse is due to the slight changes in the central frequency of the pumps that can occur between experiments. The pulse also shows some positive chirp due to the nonlinear phase imposed by the pumps. This is more clear in Fig. \ref{fig:xfrog_all}, which shows the calculated PG-XFROG traces for the synthesized field of Fig. \ref{fig:time} assuming gate pulses of $20$ and $80\ {\rm fs}$, that provide more detailed information in the temporal and spectral domains, respectively \cite{Crespo2009,Weigand2009}. The right trace ($80\ {\rm fs}$ gate) reproduces the conditions of \cite{Crespo2009,Weigand2009}, and is a good match to the experiment. The shorter gate pulse should enable the direct observation of fine temporal structure of the synthesized pulse train, provided that the relative phase between the pump pulses is kept constant throughout the measurements \cite{Weigand2009}.

 In conclusion, we developed a theoretical model capable of accurately describing the main features of highly nondegenerate CFWM of noncollinear femtosecond pulses in the spatial, spectral and temporal domains. A nonlinear propagation equation in two spatial dimensions was introduced, without imposing artificial restrictions on either the number or frequencies of the generated orders, and presented a numerical method to solve it efficiently. The simulations show that the instantaneous Kerr response and self-steepening are necessary and sufficient to reproduce the experiments and demonstrate compression without complex amplitude or phase control. They may provide a tool to study new configurations and materials for generating and compressing pulses in previously hard to access domains.

J.L.S. acknowledges financial support under Grant No. SFRH/BD/24523/2005 from FCT-MCTES, Portugal.


\end{document}